\documentclass[journal]{IEEEtran}

\usepackage{graphicx,amssymb,lineno}
\usepackage{amsmath,amsfonts,amssymb}

\usepackage{algorithmic}
\usepackage[usenames]{color}
\usepackage{float}

\usepackage[linesnumbered,ruled,vlined]{algorithm2e}

\usepackage{graphicx,graphics,color,epsfig,subfigure,graphpap,rotate}
\usepackage{times, verbatim, subfigure, epsfig, graphicx, latexsym, amsmath}
\usepackage{url}
\usepackage{subfigure}
\begin{document}
%

\title{Energy Efficient Scheduling for mmWave Backhauling of Small Cells in Heterogeneous Cellular Networks}

\author{Yong~Niu,
         Chuhan~Gao,
         Yong~Li,~\IEEEmembership{Member,~IEEE,}
         Li~Su,
         Depeng~Jin,~\IEEEmembership{Member,~IEEE}
\thanks{Y. Niu, C. Gao, Y. Li, L.~Su and D.~Jin are with State Key Laboratory on
 Microwave and Digital Communications, Tsinghua National Laboratory for Information
 Science and Technology (TNLIST), Department of Electronic Engineering, Tsinghua
 University, Beijing 100084, China (E-mails: liyong07@tsinghua.edu.cn).} 
}%

\maketitle

\begin{abstract}

Heterogeneous cellular networks with small cells densely deployed underlying the conventional homogeneous macrocells are emerging as a promising candidate for the fifth generation (5G) mobile network. When a large number of base stations are deployed, the cost-effective, flexible, and green backhaul solution becomes one of the most urgent and critical challenges. With vast amounts of spectrum available, wireless backhaul in the millimeter wave (mmWave) band is able to provide several-Gbps transmission rates. To overcome high propagation loss at higher frequencies, mmWave backhaul utilize beamforming to achieve directional transmission, and concurrent transmissions (spatial reuse) under low inter-link interference can be enabled to significantly improve network capacity. To achieve an energy efficient solution for the mmWave backhauling of small cells, we first formulate the problem of minimizing the energy consumption via concurrent transmission scheduling and power control into a mixed integer nonlinear programming problem. Then we develop an energy efficient and practical mmWave backhauling scheme, where the maximum independent set based scheduling algorithm and the power control algorithm are proposed to exploit the spatial reuse for low energy consumption and high energy efficiency. We also theoretically analyze the conditions that our scheme reduces energy consumption, and the choice of the interference threshold for energy reduction. Through extensive simulations under various traffic patterns and system parameters, we demonstrate the superior performance of our scheme in terms of energy consumption and energy efficiency, and also analyze the choice of the interference threshold under different traffic loads, BS distributions, and the maximum transmission power.

\end{abstract}

\section{Introduction}\label{S1}

The volume of mobile traffic is exploding, and some industry and academic experts predict a 1000-fold demand increase by 2020 \cite{data1, data2}. In order to achieve 1000x increase in the offered data rates and throughput with respect to current state-of-the-art technology, heterogeneous cellular networks (HCNs) with small cells densely deployed underlying the conventional homogeneous macrocells emerge as a promising candidate for the fifth generation (5G) mobile network \cite{inter_digital}. In the 5G era, with a large number of base stations deployed, the cost-effective,
flexible, and green backhaul solution becomes one of the most urgent and critical challenges in the design of HCNs. Since it is costly, inflexible, and time-consuming to connect the densely deployed small cells by wired fiber-based backhaul, operators are estimating that 80\%
of the small cells will be connected with wireless backhaul in 5G \cite{inter_digital}. With vast amounts of spectrum available, wireless backhaul in the millimeter wave (mmWave) band, such as the 60 GHz band and E-band (71--76 GHz and 81--86 GHz), is able to provide several-Gbps data rates, and has gained considerable interest recently \cite{backhaul_5G}. Furthermore, rapid development of integrated circuits and antennas in the mmWave band paves the way for mmWave communications to make a big impact in 5G \cite{CMOS,CMOS2,CMOS3}. Several standards have been defined for indoor wireless personal area networks (WPAN) or wireless local area networks (WLAN) in the mmWave band, such as ECMA-387 \cite{ECMA 387}, IEEE 802.15.3c \cite{IEEE 802.15.3c}, and IEEE 802.11ad
\cite{IEEE 802.11ad}.

\begin{figure} [htbp] 
\begin{center}
\includegraphics*[width=9cm]{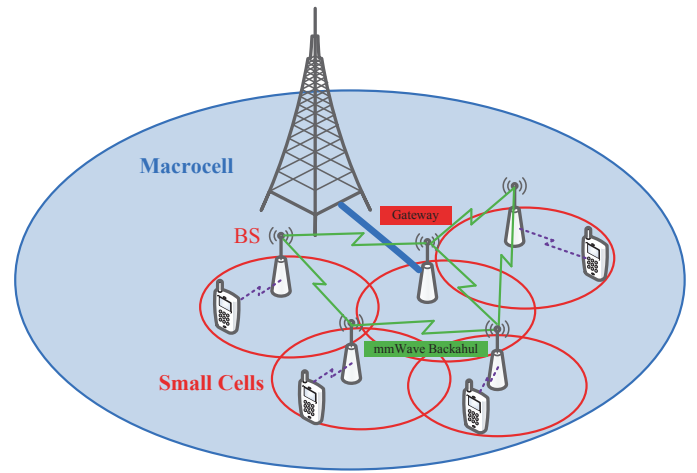}
\end{center}
\caption{The mmWave mesh backhaul network for the small cells densely deployed underlying the macrocell.}
\label{small cell}
\end{figure}

In Fig. \ref{small cell}, we show a typical scenario of the mmWave mesh backhaul network for the small cells densely deployed. Mobile users are associated with base stations (BSs) of the small cells, and BSs are connected by mesh backhaul network in the mmWave band. There are one or more BSs connected to the Internet via the macrocell, which are called gateways. With the access links of small cells also in the mmWave band to achieve high access rates, the cell radius becomes smaller, and the ultra dense deployment of small cells is anticipated for the 5G networks \cite{ultra_dense}. With small cells densely deployed, the energy consumption will increase tremendously. In order to keep the energy consumption at today's level, the energy efficient backhaul scheduling design is required \cite{backhaul_5G}.

mmWave communications have unique characteristics that are different from existing communication systems at lower carrier frequency. Due to the high carrier frequency, mmWave communications suffer from high propagation loss. For example, the free space
path loss at the 60 GHz band is 28 decibels (dB) more than that at 2.4 GHz \cite{singh_outdoor}. Consequently, directional antennas are synthesized at both the transmitter and receiver to achieve high antenna gain by the beamforming technique \cite{beam_training, Beamtraining2, beam_training2}. With a small wavelength, low-cost and compact on-chip and in-package directional antennas can be synthesized in a small platform \cite{CMOS3}. In this case, the omnidirectional carrier sensing as in WiFi is disabled, and the third party nodes are deaf to the current communication, which is referred to as the deafness problem. On the other hand, reduced interference between links enables concurrent transmissions (spatial reuse), and the network capacity can be improved significantly. To provide efficient and green backhauling for small cells densely deployed, spatial reuse should be fully exploited in the scheduling scheme design. With the enabled concurrent transmissions, there will be more time resources that can be allocated to data flows, and transmission power of flows can be reduced while each flow achieves the same or higher throughput. When the transmission time doubled, the transmission rate can be reduced by half to achieve the same throughput. Since the transmission power $P_t$ is proportional to $({2^{R/W}} - 1)$ according to the Shannon's
channel capacity ($R$ denotes the transmission rate and $W$ denotes the bandwidth), the transmission power can be reduced by more than half under relatively high SNR. Therefore, the consumed energy can be reduced while ensuring the throughput of flows.


In this paper, we develop an energy efficient mmWave backhauling scheme for small cells in 5G, where the concurrent transmissions are exploited for lower energy consumption and higher energy efficiency. Our scheme is based on the time division multiple access (TDMA). By enabling concurrent transmissions with the required throughput of each flow, our scheme allocates more time resources to each flow, and consequently the energy consumption is dramatically reduced. To the best of our knowledge, we are the first to exploit the concurrent transmissions to achieve low energy consumption and high energy efficiency for the mmWave backhaul network of small cells in HCNs. The contributions of this paper are four-fold, which are summarized as follows.


\begin{itemize}
\item We formulate the joint optimization problem of concurrent transmission scheduling and power control into a mixed integer nonlinear programming problem, i.e., minimizing the energy consumption by fully exploiting the spatial reuse with the achieved throughput of each flow not less than that under serial TDMA.
\item We propose an energy efficient mmWave backhauling scheme, which consists of maximum independent set (MIS) based scheduling algorithm and power control algorithm, to efficiently solve the formulated problem. The MIS based scheduling algorithm allocates as many flows as possible into each pairing based on the contention graph, while the power control algorithm decides the transmission duration and transmission power of each flow.
\item We give a theoretical performance analysis on the fundamental conditions that our scheme reduces the energy consumption by allocating more channel time allocations to flows, and also analyze the choice of the interference threshold for energy reduction.
\item Through extensive evaluations under various traffic patterns and system parameters, we demonstrate our scheme achieves the lowest energy consumption and the highest energy efficiency compared with other schemes. Furthermore, we investigate the choice of the interference threshold for our scheme under different traffic loads, BS distributions, and maximum transmission power.

\end{itemize}

The rest of this paper is organized as follows. Section \ref{S2} introduces the related work on the transmission scheduling for the small cells and wireless backhaul in mmWave band. Section \ref{S3} introduces the system model and illustrates our basic idea by
an example, and also formulates the joint problem of concurrent transmission scheduling and power control into a mixed integer nonlinear programming problem. Section \ref{S5} presents our proposed green backhauling scheme. In Section \ref{pre-S6}, we give a performance analysis on the conditions that our scheme reduces the energy consumption, and the choice of threshold for energy reduction. Section \ref{S6} gives the evaluation of our scheme in terms of energy consumption, network throughput, and energy efficiency compared with other schemes under various traffic patterns. Finally, we conclude this paper in Section \ref{S7}.

\section{Related Work}\label{S2}

Recently, there are some schemes proposed on the transmission scheduling of the small cells in the mmWave band. Since ECMA-387 \cite{ECMA 387} and IEEE 802.15.3c \cite{IEEE 802.15.3c} adopt TDMA, there is a class of works based on the TDMA \cite{mao_12, mao_13, Qiao_6, Qiao_15, Qiao, EX_Region, Qiao_7}. With the multi-user interference between links below a specific threshold, multiple links are scheduled to
communicate in the same slot \cite{Qiao_6, Qiao_15}. Cai \emph{et al.} \cite{EX_Region} introduced the concept of exclusive region to
support concurrent transmissions, and derived the conditions that concurrent transmissions always
outperform serial TDMA. Emphasizing the quality of service (QoS) guarantees of flows, Qiao \emph{et al.} \cite{Qiao} proposed a concurrent transmission scheduling algorithm
to maximize the number of scheduled flows with their QoS requirements satisfied. Qiao \emph{et al.} \cite{Qiao_7} also proposed a multi-hop concurrent transmission
scheme to address the link outage problem (blockage) and combat large path loss to improve
flow throughput, where the flow of low channel quality is transmitted through multiple hops of high channel quality.

There are also some centralized scheduling protocols proposed for small cells in the mmWave band \cite{Gong, MRDMAC, mao, chenqian}. Gong \emph{et al.} \cite{Gong} proposed a directional CSMA/CA protocol, where the virtual carrier sensing is exploited to overcome the deafness problem, and the network allocation vector information is distributed by the piconet coordinator. Singh \emph{et al.} \cite{MRDMAC}
proposed a multihop relay directional MAC protocol (MRDMAC), where the deafness problem is overcome
by the PNC's weighted round robin scheduling. In MRDMAC, if a wireless terminal (WT) is lost due to
blockage, the access point will choose a live WT to act as a relay to the lost WT. Son \emph{et al.}
\cite{mao} proposed a frame based directional MAC protocol (FDMAC), which achieves high efficiency by amortizing the
scheduling overhead over multiple concurrent transmissions in a row.
The core of FDMAC is the Greedy Coloring algorithm, which fully
exploits spatial reuse and greatly improves the network throughput
compared with MRDMAC \cite{MRDMAC} and memory-guided directional MAC
(MDMAC) \cite{MDMAC}. Chen \emph{et al.}
\cite{chenqian} proposed a directional cooperative MAC protocol, D-CoopMAC, to coordinate the
uplink channel access in an IEEE 802.11ad WLAN. In D-CoopMAC, a relay is selected for each direct link, and
when the two-hop link outperforms the direct link, the latter will be replaced by the former.

There are also some works on utilizing mmWave band for wireless backhaul. Lebedev \emph{et
al.} \cite{60GHz-backhaul-1} identified advantages of the 60 GHz band for short-range mobile
backhaul by feasibility analysis and comparison with E-band. Bojic \emph{et al.} \cite{60GHz-backhaul-4} discussed the advanced wireless and optical technologies for small cell mobile backhaul with dynamic software-defined management in 60 GHz and E-band. Bernardos \emph{et al.} \cite{joint_design} discussed the main challenges to design the
backhaul and radio access network jointly in a cloud-based mobile network, involving
the physical, MAC, and network layers. Islam \emph{et
al.} \cite{60GHz-backhaul-2} performed a joint cost optimal aggregator node placement, power
allocation, channel scheduling and routing to optimize the mmWave wireless backhaul network. Ge \emph{et al.} \cite{backhaul_5G} study how to
promote 5G wireless backhaul networks in high throughput and
low energy consumption ways via two typical small cell scenarios adopting millimeter wave communication technologies, i.e., the central scenario and the distribution scenario. It is shown that the distribution solution has higher energy efficiency than the central solution in 5G wireless backhaul networks. Taori \emph{et al.} \cite{2_7} proposed to multiplex the backhaul and access on the same frequency band to obtain a cost-effective and scalable wireless backhaul
solution. They also proposed a time-division multiplexing (TDM) based scheduling scheme to support point-to-multipoint, non-line-of-sight, mmWave backhaul. Recently, Niu \emph{et al.} \cite{JSAC_own} proposed a
joint scheduling scheme for the radio access and
backhaul of small cells in the 60 GHz band, where
a path selection criterion is designed to enable the device-to-device communications.

To the best of our knowledge, most of these works do not focus on energy consumption reduction problem of mmWave wireless backhaul network for small cells densely deployed. In this paper, we exploit the concurrent transmissions from the directional communications in mmWave band for the energy consumption reduction in 5G HCNs.


\section{System Overview and Problem Formulation}\label{S3}

\subsection{System Model}\label{S3-1}

\begin{figure} [htbp]
\begin{center}
\includegraphics*[width=7.5cm]{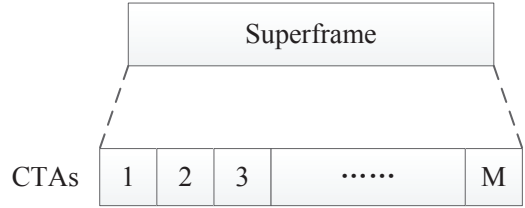}
\end{center}
\caption{The mmWave superframe structure for the backhaul network of small cells.}
\label{superframe}
\end{figure}

We consider a typical HCN shown in Fig. \ref{small cell}, where the assumed macrocell is coupled with small cells to some extent. As in \cite{mmW_5G_Qiao}, each BS has the communication modes of both 4G access and mmWave backhauling operation. We assume each BS in the backhaul network is equipped with the electronically steerable directional antenna, which enables directional transmission towards other BSs in mmWave band. The macrocell BS and small cell BSs are also equipped with omnidirectional antennas for 4G communications \cite{mmW_5G_Qiao}. In our investigated system, time is partitioned into superframes, and each superframe consists of $M$ time slots called channel time allocation (CTA), which is illustrated in Fig. \ref{superframe}. We further assume the transmission requests and signaling information for mmWave backhauling are collected by the 4G BS by its reliable transmission \cite{mmW_5G_Qiao}.

With the fixed and static BSs, the backhaul network topology and location information of BSs can be obtained. With the appropriately deployed BSs, the backhaul network mainly relies on the line-of-sight (LOS) transmissions. In the serial TDMA scheme, we assume there are $N$ flows requesting transmission time in the superframe. Since each CTA can only be occupied by one flow, we have $N \le M$. For flow $i$, we denote its sender and receiver by $s_i$ and $r_i$, respectively. The distance between $s_i$ and $r_j$ is denoted by $l_{ij}$. We also denote the antenna gain of $s_i$ in the direction of $s_i \to r_j$ by ${G_t}(i,j)$, and the antenna gain of $r_i$ in the direction of $s_j \to r_i$ by ${G_r}(j,i)$. According to the path loss model \cite{EX_Region}, the received power at $r_i$ from $s_i$ can be calculated as

\begin{equation}
{P_r}(i,i) = {k_0}{G_t}(i,i){G_r}(i,i)l_{ii}^{ - n}{P_t},
\end{equation}
where ${k_0}$ is a constant coefficient and proportional to ${(\frac{\lambda }{{4\pi }})^2}$ ($\lambda $ denotes the wavelength), $n$ is the path loss exponent, and ${P_t}$ is the transmission power \cite{Qiao, EX_Region}. Then in the serial TDMA scheme, the received SNR of flow $i$ can be calculated as

\begin{equation}
{\rm{SN}}{{\rm{R}}_i} = \frac{{{k_0}{G_t}(i,i){G_r}(i,i)l_{ii}^{ - n}{P_t}}}{{{N_0}W  }},
\end{equation}
where $W$ is the bandwidth, and ${{N_0}}$ is the onesided power spectral density of white Gaussian noise \cite{Qiao}. According to Shannon's
channel capacity, the achievable data rate of flow $i$ can be estimated as
\begin{equation}
{R_i} = \eta W{\log _2}(1 + \frac{{{k_0}{G_t}(i,i){G_r}(i,i)l_{ii}^{ - n}{P_t}}}{{{N_0}W  }}),
\end{equation}
where $\eta  \in (0,1)$ describes the efficiency of the transceiver design. If the number of CTAs allocated to flow $i$ in the serial TDMA scheme is ${\delta _i}$, the achieved throughput of flow $i$ can be expressed as

\begin{equation}\label{qi}
{q_i} = \frac{{{R_i} \cdot {\delta _i} }}{{ M }}.
\end{equation}

On the other hand, the reduced interference enables concurrent transmissions. Due to the half-duplex constraint, adjacent links, which share common vertices, cannot be scheduled for concurrent transmissions \cite{mao}. For nonadjacent flows $i$ and $j$, the received interference at $r_i$ from $s_j$ under concurrent transmissions can be calculated as

\begin{equation}
{P_r}(j,i) = \rho {k_0}{G_t}(j,i){G_r}(j,i)l_{ji}^{ - n}{P_t},
\end{equation}
where $\rho$ denotes the multi-user interference (MUI) factor related to the cross correlation of signals from different links \cite{Qiao}. Thus, the received SINR at $r_i$ can be expressed as

\begin{equation}
{\rm{SIN}}{{\rm{R}}_i} = \frac{{{k_0}{G_t}(i,i){G_r}(i,i)l_{ii}^{ - n}{P_t}}}{{{N_0}W + \rho \sum\limits_{j } {{k_0}{G_t}(j,i){G_r}(j,i)l_{ji}^{ - n}{P_t}} }}.
\end{equation}
Similarly, the achieved transmission rate in this case can be obtained according to the Shannon's
channel capacity.

\subsection{Motivation and Main Idea}\label{S3-2}

To reduce the energy consumption with the throughput of flows ensured, we should enable concurrent transmissions of flows to allocate more CTAs to each flow.

\begin{figure} [htbp]
\begin{center}
\includegraphics*[width=8cm]{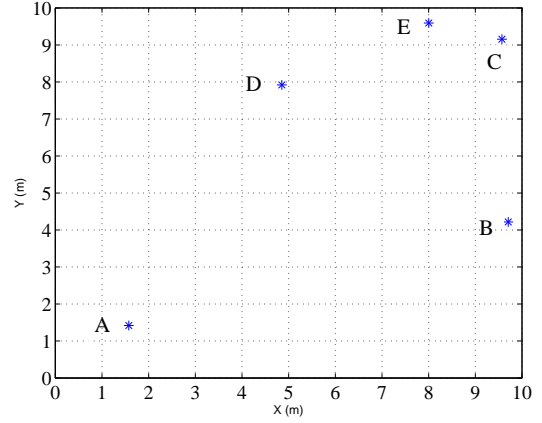}
\end{center}
\caption{The network topology of five nodes in the example.}
\label{topology}
\end{figure}

\begin{figure} [htbp]
\begin{center}
\includegraphics*[width=8cm]{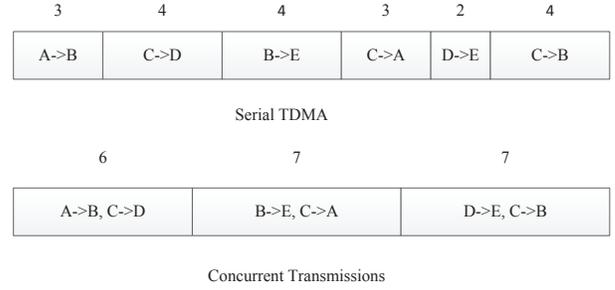}
\end{center}
\caption{The CTA allocation under the serial TDMA scheme and with concurrent transmissions.}
\label{example}
\end{figure}

We utilize a typical example to illustrate our basic idea. There are five nodes uniformly distributed in a square area of 10 $ \times$ 10 $m^2$, and they are denoted by $A$, $B$, $C$, $D$, and $E$, respectively, as shown in Fig. \ref{topology}. There are 20 CTAs in the superframe, and six flows in the network, i.e., flow $A \to B$, flow $C \to D$, flow $B \to E$, flow $C \to A$, flow $D \to E$, flow $C \to B$. In the serial TDMA scheme, there are 3, 4, 4, 3, 2, and 4 CTAs allocated to flow $A \to B$, flow $C \to D$, flow $B \to E$, flow $C \to A$, flow $D \to E$, and flow $C \to B$, respectively. By exploiting concurrent transmissions, we can allocate more CTAs to each flow. As shown in Fig. \ref{example}, flow $A \to B$ and flow $C \to D$ are scheduled for concurrent transmissions, and 6 CTAs are allocated to them. There are 7 CTAs allocated to flow $B \to E$ and flow $C \to A$, and 7 CTAs allocated to flow $D \to E$ and flow $C \to B$. For simplicity to illustrate this example, we assume the interference between nonadjacent flows can be neglected \cite{mao}. Since more CTAs are allocated to each flow, the transmission rate and transmission power can be reduced to achieve the same throughput of serial TDMA scheme. For example, adopting the simulation parameters in \cite{Qiao}, the transmission power of flow $A \to B$ can be reduced from 0.1 mW to about 0.03 mW. Although the transmission time is twice of that in the serial TDMA scheme, the transmission power is reduced by more than half. Thus, the energy consumption of flow $A \to B$ can be reduced. The total energy consumption of all flows can be reduced by about 48.8\% for the example. Thus, the energy consumption can be reduced significantly by allocating more CTAs to each flow.

From the above example, we can observe that the concurrent transmission scheduling and transmission power allocation after scheduling are key mechanisms to reduce the energy consumption, and should be optimized to achieve low energy consumption.

\subsection{Problem Formulation}\label{S4}

If concurrent transmissions are enabled, there will be more CTAs allocated to each flow, and the transmission power of flows can be reduced with the achieved throughput of each flow not less than that under serial TDMA, which reduces the energy consumption eventually. In this subsection, we formulate the joint optimization problem of concurrent transmission scheduling and power control to minimize energy consumption.


We denote the schedule for the superframe by $\textbf{S}$, and assume it has $K$ pairings, where a pairing is defined as several continuous CTAs. In each pairing, the same group of links are scheduled for concurrent transmissions \cite{mao}. For each flow $i$, we define a binary variable $a_i^k$ to indicate whether flow $i$ is scheduled in the $k$th pairing. If it is, $a_i^k=1$; otherwise, $a_i^k=0$. The number of CTAs in the $k$th pairing is denoted by $\theta^k$. To simplify the power control mechanism, we assume the transmission power of each flow is fixed during the superframe, and the transmission of each flow is completed in one pairing. With the transmission power of flow $i$ denoted by $P_t^i$, we can obtain the transmission rate of flow $i$ in the $k$th pairing, denoted by $R^k_i$, as

\begin{equation}
R_i^k = \eta W{\log _2}(1 + \frac{{a_i^k{k_0}{G_t}(i,i){G_r}(i,i)l_{ii}^{ - n}{P_t^i}}}{{{N_0}W + \rho \sum\limits_{j} {a_j^k{k_0}{G_t}(j,i){G_r}(j,i)l_{ji}^{ - n}{P_t^j}} }}).
\end{equation}

To reduce energy consumption and also ensure efficiency, we should minimize the energy consumption of flows while achieving the required throughput of each flow. The total consumed energy can be expressed as

\begin{equation}
\sum\limits_{i = 1}^N {\sum\limits_{k = 1}^K {P_t^i \cdot a_i^k \cdot {\theta ^k} \cdot \Delta T} },
\end{equation}
where $\Delta T$ denotes the time duration of each CTA. Since $\Delta T$ is constant, the objective function to be minimized can be simplified as
\begin{equation}
\sum\limits_{i = 1}^N {\sum\limits_{k = 1}^K {P_t^i \cdot a_i^k \cdot {\theta ^k} } }.
\end{equation}

Now, we analyze the system constraints of this optimization problem. First, each flow can be only scheduled in one pairing of the superframe, which can be expressed as

\begin{equation}\label{cons1}
\sum\limits_{k = 1}^K {a_i^k = 1},\;\;\;\forall \;i.
\end{equation}

Second, there are at most $M$ CTAs in the superframe, which indicates

\begin{equation}\label{cons2}
\sum\limits_{k = 1}^K {{\theta ^k}}  \le M.
\end{equation}

Third, adjacent links cannot be scheduled in the same pairing for concurrent transmissions, which can be expressed as

\begin{equation}\label{cons3}
a_i^k + a_j^k \le 1,\;{\rm{if}}\;{\rm{flow}}\;i\;{\rm{and}}\;j\;{\rm{are}}\;{\rm{adjacent,}}\;\forall \;i,j,k.
\end{equation}

Fourth, the achieved throughput of each flow should be larger than or equal to the flow transmission requirement, which can be expressed as follows,

\begin{equation}\label{cons4}
\frac{{\sum\limits_{k = 1}^K {R_i^k \cdot {\theta ^k}} }}{M} \ge {q_i},\;\forall \;i,
\end{equation}
where $q_i$ is defined in (\ref{qi}), and the left side represents the achieved throughput of flow $i$.

Finally, the transmission power of each flow should not exceed the transmission power in the serial TDMA scheme, which can be expressed as follows.     \begin{equation}\label{cons5}
P_t^i \le {P_t},\;\forall \;i.
\end{equation}

Therefore, the joint problem of concurrent transmission scheduling and power control to minimize energy consumption can be formulated as follows.

\begin{equation}\hspace{-1.25cm}
({\rm{P1}})\ \ \min \sum\limits_{i = 1}^N {\sum\limits_{k = 1}^K {P_t^i \cdot a_i^k \cdot {\theta ^k} } },\label{OBJ}
\end{equation}

\hspace{2.2cm}s. t.
\hspace{0.2cm}Constraints (\ref{cons1})--(\ref{cons5}).

This is a mixed integer nonlinear programming (MINLP) problem, where $P_t^i$ is the continuous variable, $a_i^k$ is the binary variable, and $\theta ^k$ is the integer variable. The transmission power of each flow cannot be determined until the scheduling decision for this pairing is made because of the interference between concurrent flows. This problem is more complex than the 0--1 Knapsack problem, which is NP-complete \cite{Knapsack, Qiao}. In the next section, we propose a heuristic concurrent transmission scheduling and a power control algorithm to solve problem P1 with low complexity.



\section{Scheduling and Power Control Algorithms}\label{S5}

The problem of P1 has two key parts to be optimized, the concurrent transmission scheduling including the CTA allocation, and the transmission power control of flows in each pairing. To achieve a practical solution, these two parts are completed separately. In this section, we propose a heuristic concurrent transmission scheduling and power control algorithm for the formulated problem, which fully exploits the concurrent transmissions to reduce the energy consumption. First, we model the contention relationship among flows by the contention graph. Based on it, we allocate the flows into different pairings by the minimum degree maximum independent set (MIS) algorithm. Then, we propose a power control algorithm to adjust the transmission power of flows for lower energy consumption with the achieved throughput of each flow ensured.

\subsection{Contention Graph}\label{S5-1}

The contention graph illustrates the contention relationship among flows as in \cite{yun_1}. In the contention graph, each vertex represents one flow in the network, and there is one edge between two vertices if there is severe interference between these two flows. For flow $i$ and flow $j$, we define the weight as the maximum of the interference between them, which can be expressed as

\begin{equation}
{W_{ij}} = \max \{  {P_r}(j,i), {P_r}(i,j)\}.
\end{equation}
The transmission power here is the same as that in the serial TDMA scheme, $P_t$. Then we define the interference threshold, $\sigma$, to construct the contention graph. The contention graph is constructed in the way that if the weight between two vertices normalized by $P_t$ is less than the threshold, i.e., if ${{{W_{ij}}}/{{{P_t}}}} < \sigma $ for flow $i$ and $j$, there will be no edge between these two vertices. Otherwise, there will be an edge between them. Of course, for adjacent flows, there will always be an edge between them since adjacent flows cannot be scheduled for concurrent transmissions due to the half-duplex assumption.

\begin{figure} [htbp]
\begin{center}
\includegraphics*[width=3cm]{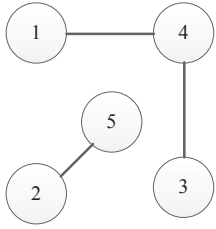}
\end{center}
\caption{An example of the contention graph.}
\label{contention graph}
\end{figure}

In Fig. \ref{contention graph}, we show an example of the contention graph. We can observe that there is an edge between flow 2 and flow 5, which indicates there is severe interference between flow 2 and flow 5, or flow 2 and flow 5 are adjacent.

\subsection{MIS based Scheduling Algorithm}\label{S5-2}

Based on the contention graph, we propose a maximum independent set (MIS) based scheduling algorithm to allocate flows into different pairings. To fully exploit concurrent transmissions, we utilize the maximum independent set to enable as many concurrent transmissions as possible in each pairing. The maximum independent set of the contention graph is a set of flows that have no edge between each other on the contention graph with the maximum cardinality \cite{yun_1}. For the example in Fig. \ref{contention graph}, the set of flow 1, 2, and 3, $\{ 1,2,3\} $, is a maximum independent set of the contention graph. Since obtaining the MIS of a general graph is NP-complete, we adopt the minimum-degree greedy algorithm to approximate the maximum
independent set \cite{yun_1_15}. This algorithm is proved to achieve a performance ratio of $(\Delta  + 2)/3$ for the graphs with degree
bounded by $\Delta$. The MIS based scheduling algorithm iteratively allocates each flow into each pairing by the minimum-degree greedy algorithm until all flows are allocated.


We denote the contention graph by $G(V,E)$, where $V$ denotes the set of vertices in the contention graph, and $E$ denotes the set of edges in the contention graph. We refer to two vertices as neighbors if there is one edge between them. For any vertex $v \in V$, we denote the set of its neighboring vertices by $N(v)$. For any vertex $v \in V$, we denote its degree by $d(v)$. In graph theory, the degree of a vertex of a graph is the number of edges incident to the vertex. We denote the set of flows scheduled in the $k$th pairing by $V^k$, and $V_u^k$ denotes the set of unvisited and possible flows to be scheduled in the $k$th pairing.

The pseudo-code of the MIS based scheduling algorithm is presented in Algorithm \ref{alg:MIS}. The algorithm iteratively schedules flows into each pairing until all flows are scheduled, as indicated by line 3. In each pairing, $V_u^k$ is initialized to the set of unscheduled flows, $V$, in line 6, and flows are scheduled into each pairing by the minimum-degree greedy algorithm, as indicated by lines 7--10. In line 11, the scheduled flows of the $k$th pairing are removed from $V$. The computational complexity of Algorithm \ref{alg:MIS} is $\mathcal{O}(|V|^2)$.

\begin{algorithm}[htbp]
 \DontPrintSemicolon
 \caption{MIS Based Concurrent Transmission Scheduling.}\label{alg:MIS}
  \textbf{Input:} The contention graph, $G(V,E)$; \\

  \textbf{Initialization:}  $k$=0;\\
    \While {$|V| > 0$}
    {
        $k$=$k$+1;  \\
        Set ${V^k} = \emptyset $; \\Set $V_u^k$ with $V_u^k=V$; \\
        \While {$|{V_u^k}|>0$ }
        {
            Obtain $v\in V_u^k$ such that $d(v) = \mathop {\min }\limits_{w \in V_u^k} d(w)$; \\
            $V^k=V^k \cup v$;\\
            $V_u^k = V_u^k - \{{v}\cup N(v)\}$;\\
        }
    $V=V-V^k$;\\
    }

$\mathbf{Return}\ V^k$ for each pairing.
\end{algorithm}

\subsection{Power Control Algorithm} \label{S5-3}

With the flows allocated into each pairing by the MIS based scheduling algorithm, we propose a power control algorithm to decide the duration of each pairing and the transmission power of each flow. Our algorithm aims to reduce the energy consumption of flows by allocating more CTAs to flows compared with the serial TDMA scheme.

For each flow $i$ in the $k$th pairing, we can obtain its transmission rate with the transmission power of each flow same as that in the serial TDMA scheme, which can be expressed as

\begin{equation}
R_i' = \eta W{\log _2}(1 \hspace{-0.1cm}+\hspace{-0.1cm} \frac{{{k_0}{G_t}(i,i){G_r}(i,i)l_{ii}^{ - n}{P_t}}}{{{N_0}W + \rho\hspace{-0.3cm} \sum\limits_{j \in {V^k{\backslash \{ i\}}}} \hspace{-0.3cm}{{k_0}{G_t}(j,i){G_r}(j,i)l_{ji}^{ - n}{P_t}} }}).
\end{equation}
Then, we can obtain the number of CTAs needed for it to achieve the throughput in the serial TDMA scheme, $T^k_i$, which can be calculated as
\begin{equation}
T_i^k = \frac{{{q_i} \cdot M}}{{R_i'}}.
\end{equation}
The maximum of these needed CTAs for the $k$th pairing, $T^k$, can be expressed as ${T^k} = \mathop {\max \{ }\limits_{i \in {V^k}} T_i^k\}$. Then the $M$ CTAs are distributed proportionally to each pairing according to its corresponding maximum, $T^k$. We also denote the number of CTAs distributed to the $k$th pairing by $\theta^k$, which can be calculated as

\begin{equation}
{\theta ^k} = \left\lfloor {\frac{{{T^k}}}{{\sum\limits_k {{T^k}} }} \cdot M} \right\rfloor,\label{condition}
\end{equation}
and the floor operation is performed on each pairing except the final pairing, which obtains the remaining CTAs.

With the number of CTAs of each pairing decided, we are able to reduce the transmission power of each flow by activating the transmission of each flow during the whole duration of its corresponding pairing. Since ${\theta ^k} \ge {T^k} \ge T_i^k \ge \delta_i$ generally with the interference between links kept low by the threshold $\sigma$, we can allocate more CTAs to each flow than the serial TDMA scheme. For flow $i$ scheduled in the $k$th pairing, we can obtain its lowest transmission rate needed to achieve its throughput in the serial TDMA scheme as

\begin{equation}
R_i'' = \frac{{{q_i} \cdot M}}{{\theta^k}}.\label{Assume_R}
\end{equation}
Then assuming the transmission power of other flows in this pairing is the same as that in the serial TDMA scheme, $P_t$, we can obtain the transmission power needed for flow $i$ to achieve the transmission rate of $R_i''$ as

\begin{equation}
P_t^i = \hspace{-0.1cm}\frac{{({2^{\frac{{R_i''}}{{\eta W}}}} \hspace{-0.1cm}- \hspace{-0.1cm}1)({N_0}W \hspace{-0.1cm}+\hspace{-0.1cm} \rho \hspace{-0.4cm}\sum\limits_{j \in {V^k{\backslash \{ i\} }}}\hspace{-0.4cm} {{k_0}{G_t}(j,i){G_r}(j,i)l_{ji}^{ - n}{P_t}} )}}{{{k_0}{G_t}(i,i){G_r}(i,i)l_{ii}^{ - n}}}. \label{P_final}
\end{equation}

The pseudo-code of the power control algorithm is presented in Algorithm \ref{alg:PC}. The algorithm has two parts. In the first part, $T^k$ is obtained for each pairing, as indicated in lines 4--9. The second part decides the number of CTAs allocated to each pairing, and the transmission power of each flow in this pairing, as indicated in lines 10--16. Since each flow can only be scheduled into one pairing, the computational complexity of the algorithm is $\mathcal{O}(|V|)$.

\begin{algorithm}[htbp]
 \DontPrintSemicolon
 \caption{Power Control Algorithm.}\label{alg:PC}
  \textbf{Input:} The set of scheduled flows for each pairing, $V^k$; \\
        \hspace{0.95cm} The number of pairings, $K$;\\
  \textbf{Initialization:}  $k$=0;\\
    \While {$k<K$}
    {
        $k$=$k$+1;  \\
        \For {{\rm{each flow}} $i \in V^k$}
        {
            Obtain its transmission rate without power control, $R_i'$; \\
            Obtain the number of CTAs needed to achieve the required throughput by $R_i'$, $T_i^k$;\\
        }
    Obtain ${T^k} = \mathop {\max \{ }\limits_{i \in {V^k}} T_i^k\}$;\\
    }
   \hspace{0.4cm} $k$=0;\\
    \While {$k<K$}
    {
        $k$=$k$+1;  \\
        Obtain the number of CTAs in the $k$th pairing, $\theta ^k$;\\
        \For {{\rm{each flow}} $i \in V^k$}
        {
            Obtain its lowest transmission rate to achieve the required throughput, $R_i''$;\\
            Obtain its transmission power, $P_t^i$;\\
        }
    }

$\mathbf{Return}\ \theta ^k$ for each pairing and $P_t^i$ for each flow.
\end{algorithm}

\section{Performance Analysis}\label{pre-S6}

Since our scheme reduces the energy consumption by allocating more CTAs to each flow, we should analyze why and when this approach is effective. Besides, the interference threshold decides the conditions of concurrent transmissions, and has a big impact on the energy consumption of our scheme. In this section, we analyze the conditions of allocating more CTAs to flows to reduce energy consumption in our scheme, and the choice of the threshold for energy reduction.

For flow $i$, we can obtain its transmission power from (\ref{Assume_R}) and (\ref{P_final}) as


\begin{equation}
P_t^i = \hspace{-0.1cm}\frac{{({2^{\frac{{{q_i} \cdot M}}{{\eta W \theta^k}}}} \hspace{-0.1cm}- \hspace{-0.1cm}1)({N_0}W \hspace{-0.1cm}+\hspace{-0.1cm} \rho \hspace{-0.4cm}\sum\limits_{j \in {V^k{\backslash \{ i\} }}}\hspace{-0.4cm} {{k_0}{G_t}(j,i){G_r}(j,i)l_{ji}^{ - n}{P_t}} )}}{{{k_0}{G_t}(i,i){G_r}(i,i)l_{ii}^{ - n}}}. \label{P_final-2}
\end{equation}

Then we can obtain its consumed energy, $E_i^g$, as $E_i^g=P_t^i \cdot {\theta ^k} \cdot \Delta T$. With (\ref{P_final-2}) incorporated, we can obtain $E_i^g$ as

\begin{equation}
\begin{aligned}
E_i^g = &{\frac{{({2^{\frac{{{q_i} \cdot M}}{{\eta W \theta^k}}}} \hspace{-0.1cm}- \hspace{-0.1cm}1)({N_0}W \hspace{-0.1cm}+\hspace{-0.1cm} \rho \hspace{-0.4cm}\sum\limits_{j \in {V^k{\backslash \{ i\} }}}\hspace{-0.4cm} {{k_0}{G_t}(j,i){G_r}(j,i)l_{ji}^{ - n}{P_t}} )}}{{{k_0}{G_t}(i,i){G_r}(i,i)l_{ii}^{ - n}}}}\\
&\cdot {\theta ^k} \cdot \Delta T.
\end{aligned}
\end{equation}

We can also obtain the consumed energy of flow $i$ in the serial TDMA scheme, $E_i^t$, which can be expressed as

\begin{equation}
E_i^t=P_t\cdot {\delta _i}\cdot \Delta T.
\end{equation}
Then we can obtain the energy ratio of flow $i$, $r_i$, as

\begin{equation}
\begin{aligned}
&r_i=\frac{E_i^g}{E_i^t}\\
&={\frac{{({2^{\frac{{{q_i} \cdot M}}{{\eta W \theta^k}}}} \hspace{-0.1cm}- \hspace{-0.1cm}1)({N_0}W \hspace{-0.1cm}+\hspace{-0.1cm} \rho \hspace{-0.4cm}\sum\limits_{j \in {V^k{\backslash \{ i\} }}}\hspace{-0.4cm} {{k_0}{G_t}(j,i){G_r}(j,i)l_{ji}^{ - n}{P_t}} )\cdot {\theta ^k}}}{{{k_0}{G_t}(i,i){G_r}(i,i)l_{ii}^{ - n}}P_t\cdot {\delta _i}}}.
\end{aligned}
\end{equation}
Then we define the SINR of flow $i$ without power control, ${\rm{SINR}}_{i}^c$, as

\begin{equation}
{\rm{SINR}}_{i}^c=\frac{{{k_0}{G_t}(i,i){G_r}(i,i)l_{ii}^{ - n}}P_t}{{N_0}W \hspace{-0.1cm}+\hspace{-0.1cm} \rho \hspace{-0.4cm}\sum\limits_{j \in {V^k{\backslash \{ i\} }}}\hspace{-0.4cm} {{k_0}{G_t}(j,i){G_r}(j,i)l_{ji}^{ - n}{P_t}}}.
\end{equation}
Thus, $r_i$ can be expressed as

\begin{equation}
r_i=\frac{({2^{\frac{{{q_i} \cdot M}}{{\eta W \theta^k}}}} \hspace{-0.1cm}- \hspace{-0.1cm}1)\cdot {\theta ^k}}{{\rm{SINR}}_{i}^c\cdot {\delta _i}}.
\end{equation}
Since the interference between concurrent flows is limited by the threshold, denoted by $\sigma$, we can obtain

\begin{equation}
{\rm{SINR}}_{i}^c > \frac{{{k_0}{G_t}(i,i){G_r}(i,i)l_{ii}^{ - n}}P_t}{{N_0}W \hspace{-0.1cm}+\hspace{-0.1cm} (|V^k|-1)\sigma{P_t}}.
\end{equation}
Thus, we can infer that
\begin{equation}
\begin{aligned}
r_i&<\frac{({2^{\frac{{{q_i} \cdot M}}{{\eta W \theta^k}}}} \hspace{-0.1cm}- \hspace{-0.1cm}1)\cdot {\theta ^k}\cdot({N_0}W \hspace{-0.1cm}+\hspace{-0.1cm} (|V^k|-1)\sigma{P_t})}{{k_0}{G_t}(i,i){G_r}(i,i)l_{ii}^{ - n}P_t\cdot {\delta _i}}\\
&=\frac{({2^{\frac{{{q_i} \cdot M}}{{\eta W \theta^k}}}} \hspace{-0.1cm}- \hspace{-0.1cm}1)\cdot {\theta ^k}\cdot({N_0}W \hspace{-0.1cm}+\hspace{-0.1cm} (|V^k|-1)\sigma{P_t})}{{P_r}(i,i)\cdot {\delta _i}}.
\end{aligned}\label{ratio-sigma}
\end{equation}

To reduce energy consumption, we should minimize $r_i$. By minimizing the right side of (\ref{ratio-sigma}), the energy consumption can be minimized. We denote the right side of (\ref{ratio-sigma}) by $r_i^u$. Since our scheme reduces the energy consumption by allocating more CTAs to each flow, we should investigate when the energy consumption can be reduced by increasing the number of allocated CTAs. The derivative of $r_i^u$ with respect to $\theta ^k$ can be expressed as

\begin{equation}
\frac{{dr_i^u}}{{d{\theta ^k}}} = ({2^{\frac{{{q_i} \cdot M}}{{\eta W \theta^k}}}}(1-\ln2 \cdot {\frac{{{q_i} \cdot M}}{{\eta W \theta^k}}})-1)\cdot \beta,
\end{equation}
where
\begin{equation}
\beta=\frac{({N_0}W \hspace{-0.1cm}+\hspace{-0.1cm} (|V^k|-1)\sigma{P_t})}{{P_r}(i,i)\cdot {\delta _i}}.
\end{equation}
Therefore, we can obtain a sufficient condition to make the derivative less than 0, i.e.,

\begin{equation}
1-\ln2 \cdot {\frac{{{q_i} \cdot M}}{{\eta W \theta^k}}}<0,
\end{equation}
which can be further converted to

\begin{equation}
\theta^k<\frac{{{q_i} \cdot M}}{{\eta W }} \cdot \ln2.\label{theta_sufficient}
\end{equation}

Therefore, energy consumption can be reduced when $\theta^k$ satisfies (\ref{theta_sufficient}). For example, if $q_i=2 {\rm{Gbps}}$, $M=5000$, $\eta=0.5$, and $W=2.16 {\rm{GHz}}$, the condition is $\theta^k \le 6418$ since $\theta^k$ is an integer. Considering the total number of CTAs is only 5000, the condition always holds for the example. We can observe that the energy consumption can be reduced by increasing the number of allocated CTAs in a large range, and the condition is satisfied more easily when the achieved throughput of flow $i$, $q_i$, is larger.

In the following, we analyze the choice of the threshold to minimize the energy consumption. As we can observe in (\ref{ratio-sigma}), the threshold cannot be too large, and $r_i$ should be less than 1, to reduce energy consumption. By making the right side of (\ref{ratio-sigma}) less than 1, we have

\begin{equation}
\sigma<{\frac{{P_r}(i,i)\cdot {\delta _i}}{({2^{\frac{{{q_i} \cdot M}}{{\eta W \theta^k}}}} \hspace{-0.1cm}- 1)\cdot {\theta ^k}P_t(|V^k|-1)}}-\frac{{N_0}W}{P_t(|V^k|-1)}.\label{sigma-cds}
\end{equation}
We can observe that with the increase of concurrent flows, the threshold should be smaller to reduce energy consumption. With the increase of the requested throughput of flow $i$, $q_i$, the threshold should decrease to ensure $r_i<1$. With a more number of CTAs allocated to flow $i$, the threshold can be larger. However, when $\sigma$ is too small to enable concurrent transmissions, our scheme will be reduced to the serial TDMA scheme, and $|V^k|$ will be equal to 1.

We denote the right side of (\ref{sigma-cds}) by $\alpha_i$. Then to reduce the energy consumption of each flow, $\sigma$ should satisfy

\begin{equation}
\sigma<\mathop {\min }\limits_{i \in N} {\alpha _i},
\end{equation}
where we also use $N$ to denote the set of all flows.

We can observe that the optimized choice of the interference threshold is different under different system parameters. In the next section, we will further investigate this problem by performance simulation.

\section{Performance Evaluation}\label{S6}

In this section, we evaluate the performance of our green backhauling scheme under various traffic patterns and system parameters. Specifically, we compare our scheme with two schemes in terms of energy consumption, achieved network throughput, and energy efficiency. Besides, we also investigate the impact of the threshold choice on the performance of our scheme under different traffic loads, BS distributions, and maximum transmission power.

\subsection{Simulation Setup}\label{S6-1}

In the simulation, we consider a dense deployment of small cells in the HCNs, where ten BSs are uniformly distributed in a square area of 100 $ \times$ 100 $m^2$. In the simulation, we adopt the widely used realistic directional antenna model, which is a main lobe of Gaussian form in linear scale and constant level of side lobes \cite{chen_2}. The gain of a
directional antenna in units of decibel (dB), denoted by $G(\theta )$, can be expressed as

\begin{equation}
G(\theta ) = \left\{ {\begin{array}{*{20}{c}}
{{G_0} - 3.01 \cdot {{(\frac{{2\theta }}{{{\theta _{ - 3{\rm{dB}}}}}})}^2},\;0^ \circ \le \theta  \le {\theta _{ml}}/2};\\
{{G_{sl}},\hspace{2.0cm}{\theta _{ml}}/2 \le \theta  \le {{180}^ \circ }};
\end{array}} \right.
\end{equation}
where $\theta$ denotes an arbitrary angle within the range $[0^ \circ, 180 ^ \circ]$, ${{\theta _{ - 3{\rm{dB}}}}}$ denotes the angle of the half-power beamwidth, and ${{\theta _{ml}}}$ denotes the main lobe width in units of degrees. The relationship between ${{\theta _{ml}}}$ and ${{\theta _{ - 3{\rm{dB}}}}}$ is ${\theta _{ml}} = 2.6 \cdot {\theta _{ - 3{\rm{dB}}}}$.
${{G_0}}$ is the maximum antenna gain, and can be obtained by

\begin{equation}
{G_0} = 10\log {(\frac{{1.6162}}{{\sin ({\theta _{ - 3{\rm{dB}}}}/2)}})^2}.
\end{equation}
The side lobe gain, ${{G_{sl}}}$, can be expressed as

\begin{equation}
{G_{sl}} =  - 0.4111 \cdot \ln ({\theta _{ - 3{\rm{dB}}}}) - {\rm{10}}{\rm{.579}}.
\end{equation}

The simulation parameters are summarized in Table \ref{tab:simulation_parameter}. In the simulation, we adopt two kinds of traffic modes:

\subsubsection {\textbf{Traffic A}} There are ten flows in the network, and the requested throughput of each flow is uniformly distributed between a predetermined throughput interval, $[q_{{\rm{down}}}, q_{{\rm{up}}}]$. We set five throughput intervals, i.e., $[0.5, 1.5]$ Gbps, $[1, 2]$ Gbps, $[1.5, 2.5]$ Gbps, $[2, 3]$ Gbps, $[2.5, 3.5]$ Gbps, and denote their traffic loads as 1, 2, 3, 4, and 5, respectively.

\subsubsection {\textbf{Traffic B}} The requested throughput of each flow is uniformly distributed between $[2.5, 3.5]$ Gbps, and we vary the number of flows in the network from 6 to 10.

In order to show the advantages of our scheme in terms of network consumption and throughput, we evaluate the following five performance metrics:

1) \textbf{Energy Consumption:} The total energy consumption of all flows in the network.

2) \textbf{Network Throughput:} The achieved throughput of all flows in the network.

3) \textbf{Energy Efficiency:} We define the energy efficiency as the achieved network throughput divided by the consumed energy in bit/s/J, which evaluates how the energy is consumed efficiently to achieve the network throughput.

4) \textbf{Energy Ratio:} The energy consumption of our scheme divided by that of the TDMA scheme.

5) \textbf{Throughput Ratio:} The achieved network throughput of our scheme divided by that of the TDMA scheme.

To show the advantages of our scheme in reducing energy consumption, we compare our scheme with two other schemes, the serial TDMA scheme and the concurrent transmission scheme with maximum transmission power (CTFP). In CTFP, the scheduling results are the same as those of our scheme, but the transmission power of all flows is fixed as $P_t$. To obtain representative results, all the simulation results are averaged from 50 independent experiments.

\begin{table}
\begin{center}
\caption{Simulation Parameters}
\def \temptablewidth {0.9\textwidth}
\begin{tabular}{ccc}
\hline
\textbf{Parameter}&\textbf{Symbol}&\textbf{Value}\\
\hline
System bandwidth & W & 2160 MHz \\
Background noise &$N_0$& -134dBm/MHz\\
Path loss exponent & $n$ & 2\\
Maximum Transmission power & $P_t$ & 40 dBm\\
MUI factor & $\rho$ & 0.01\\
CTA duration &$\Delta T$& 18 $\mu$s\\
Number of CTAs & $M$ & 5000\\
Half-power beamwidth & ${{\theta _{ - 3{\rm{dB}}}}}$ & ${\rm{30}}^\circ $\\
Threshold & $\sigma$ & ${10^{ - 10}}$ \\
\hline
\end{tabular}
\label{tab:simulation_parameter}
\end{center}
\end{table}

\subsection{Comparison with Other Schemes} \label{S6-2}

\subsubsection{Energy Consumption}

\begin{figure} [htbp]
\begin{center}
\includegraphics*[width=9cm]{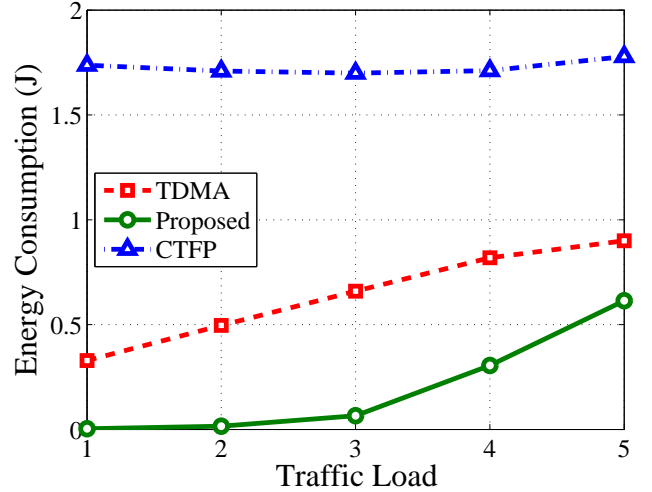}
\end{center}
\caption{The energy consumption comparison of three schemes under Traffic A.}
\label{Energy_A}
\end{figure}

In Fig. \ref{Energy_A}, we plot energy consumption comparison of the three schemes under Traffic A, where the energy consumption is in the unit of joule, denoted by J. From the results, we can observe that our scheme consumes the least energy, and the consumed energy increases with the traffic load slowly. Under light load, the gap between TDMA and our scheme is larger than that under heavy load. This is due to that under light load, there are more CTAs allocated to each flow to achieve the required throughput, and the transmission power of flows can be reduced further. At the traffic load of 5, our scheme saves about 31.8\% and 65.5\% energy compared with the TDMA scheme and CTFP, respectively. Since CTFP enables concurrent transmissions with the maximum transmission power, the gap between CTFP and our scheme is even larger.

\begin{figure} [htbp]
\begin{center}
\includegraphics*[width=9cm]{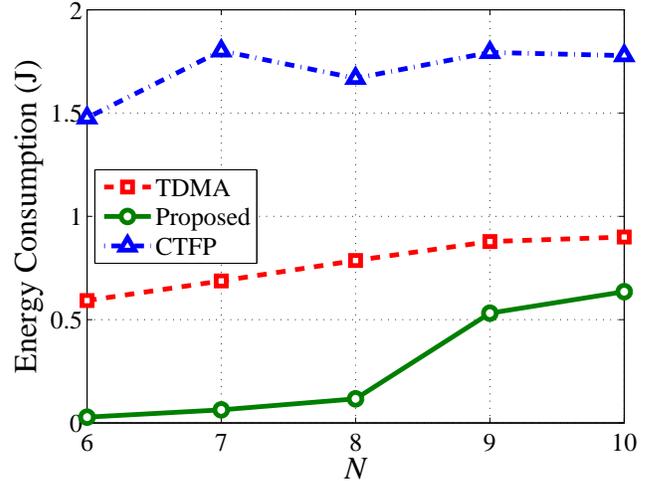}
\end{center}
\caption{The energy consumption comparison of three schemes under Traffic B.}
\label{Energy_B}
\end{figure}

In Fig. \ref{Energy_B}, we plot the energy consumption comparison of the three schemes under Traffic B. Again, our scheme achieves the lowest energy consumption among these three schemes. With the increase of the number of flows, our energy consumption increases. In TDMA scheme, the energy consumption almost increases linearly with the number of flows. The energy consumption of CTFP keeps at a high level since flows are transmitting at all CTAs with the maximum transmission power, $P_t$. In summary, our scheme consumes the lowest energy under both traffic modes compared with TDMA and CTFP with throughput of each flow ensured.

\subsubsection{Network Throughput}

\begin{figure} [htbp]
\begin{center}
\includegraphics*[width=9cm]{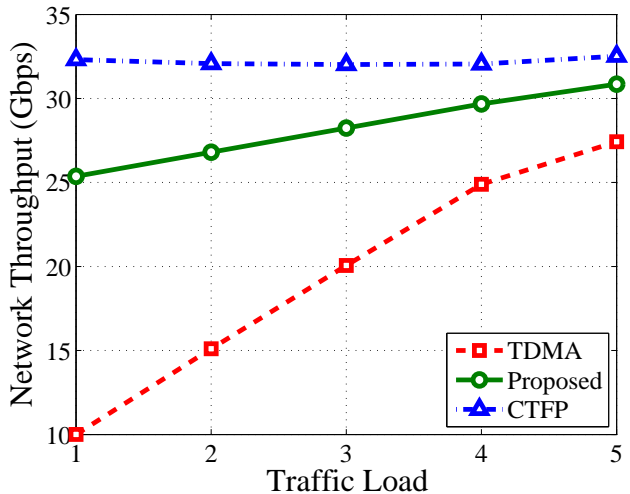}
\end{center}
\caption{The network throughput comparison of three schemes under Traffic A.}
\label{TH_A}
\end{figure}

In Fig. \ref{TH_A}, we plot the achieved network throughput of three schemes under Traffic A. We can observe that our scheme achieves higher network throughput compared with TDMA scheme. Although we assume the transmission power of other flows in the same pairing to be $P_t$ when we infer the transmission power of each flow in (\ref{P_final}), the transmission power of flows is actually lower than $P_t$ after power control, and consequently the interference power is reduced. Thus, the actual transmission rates of flows can be much higher than the assumed transmission rates, $R_i''$ in (\ref{Assume_R}). The gap between our scheme and TDMA decreases with the traffic load. This is due to under light load, there are more CTAs allocated to each flow, and the transmission power of flows can be reduced further compared with $P_t$. Since flows are transmitted at all CTAs with the maximum transmission power in the CTFP scheme, its achieved network throughput is higher than that of our scheme. With the increase of the traffic load, the gap between the CTFP scheme and our scheme decreases. At the traffic load of 5, the gap is only about 5.12\% of the network throughput of CTFP. However, our scheme consumes less energy, and has higher energy efficiency than the CTFP scheme, which will be illustrated below.


\begin{figure} [htbp]
\begin{center}
\includegraphics*[width=9cm]{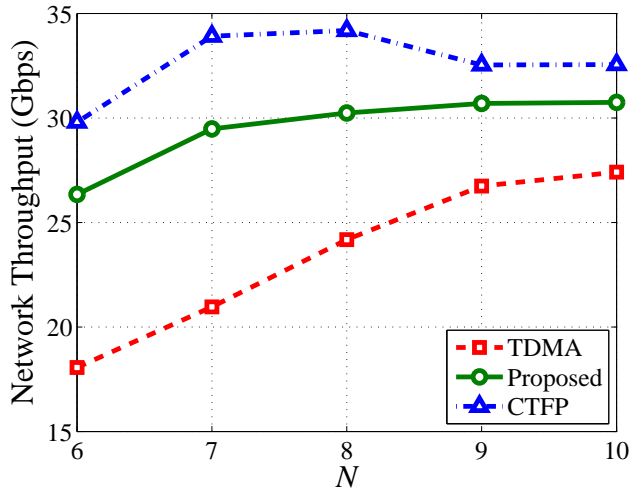}
\end{center}
\caption{The network throughput comparison of three schemes under Traffic B.}
\label{TH_B}
\end{figure}

We also plot the network throughput comparison of the three schemes under Traffic B in Fig. \ref{TH_B}. As we can observe, our scheme also has higher network throughput than the serial TDMA scheme. With large number of flows in the network (e.g., $N$ is 9 or 10), our scheme also achieves comparable performance compared with the CTFP scheme. Similar to the results in Fig. \ref{TH_A}, the gap between TDMA and our scheme is large with a small number of flows. When $N$ is small, there are more CTAs distributed to each flow, and the transmission power of flows can be reduced further. With the increase of the number of flows, the gap becomes small since the interference between flows in our scheme increases. From the results in Fig. \ref{TH_A} and Fig. \ref{TH_B}, we can observe that our scheme achieves higher throughput than TDMA, and although CTFP achieves the highest throughput, its consumed energy is much higher and energy efficiency is much lower than our scheme, which will be illustrated as follows.

\subsubsection{Energy Efficiency}

\begin{figure} [htbp]
\begin{center}
\includegraphics*[width=9cm]{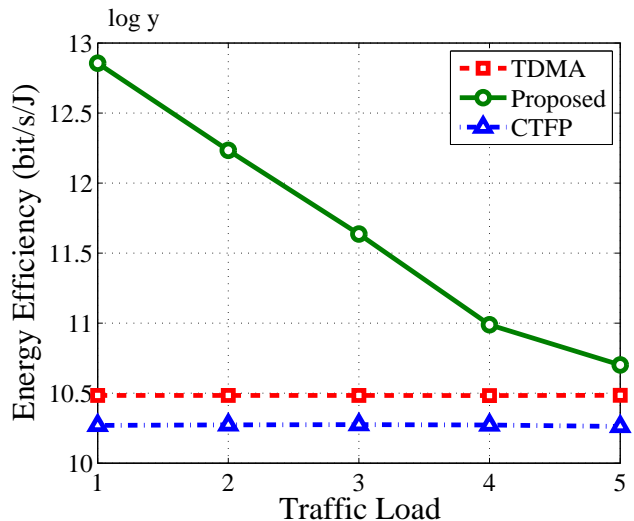}
\end{center}
\caption{The energy efficiency comparison of three schemes under Traffic A.}
\label{EE_A}
\end{figure}

In Fig. \ref{EE_A}, we plot the energy efficiency comparison of the three schemes under Traffic A. Since the energy efficiency of our scheme is very large, we show the results with y axis using the logarithmic coordinates, i.e., the values of the energy efficiency are taken after the ${\rm{log}}_{10}$ operation. From the results, we can observe that our scheme has much higher energy efficiency than the other two schemes. Our energy efficiency decreases with traffic load, which can be inferred from the results in Fig. \ref{Energy_A} and Fig. \ref{TH_A}. With the increase of traffic load, the transmission power of flows increases, and the interference between flows increases. Consequently, the benefit from concurrent transmissions becomes smaller, which decreases the energy efficiency eventually. TDMA has higher energy efficiency than the CTFP scheme since CTFP consumes much more energy while the gain from higher transmission power is limited by the interference.

\begin{figure} [htbp]
\begin{center}
\includegraphics*[width=9cm]{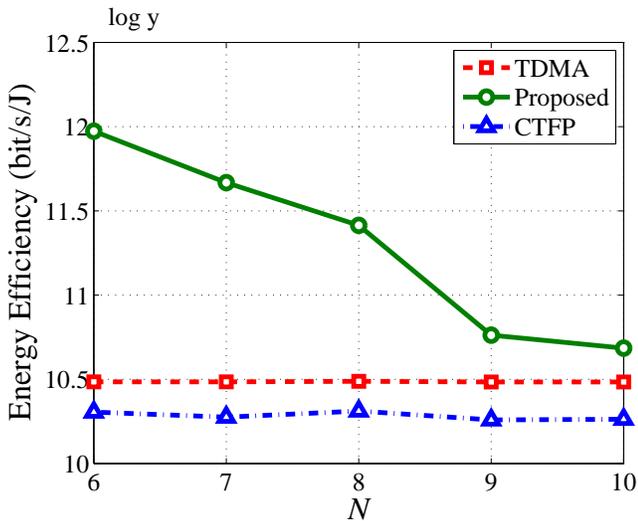}
\end{center}
\caption{The energy efficiency comparison of three schemes under Traffic B.}
\label{EE_B}
\end{figure}

We also plot the energy efficiency comparison of the three schemes under Traffic B in Fig. \ref{EE_B}. Our scheme also achieves the highest energy efficiency among the three schemes. With the increase of the number of flows, our energy efficiency decreases similar to the results in Fig. \ref{EE_A}. From the results above, we can observe that our scheme has much higher energy efficiency than TDMA and CTFP under both traffic modes.

\subsection{Choice of the Interference Threshold} \label{S6-3}

Since the choice of threshold has an important impact on the performance of our scheme, we now investigate it under different system parameters.

\subsubsection{Under Different Traffic Loads}

\begin{figure} [htbp]
\begin{center}
\includegraphics*[width=9cm]{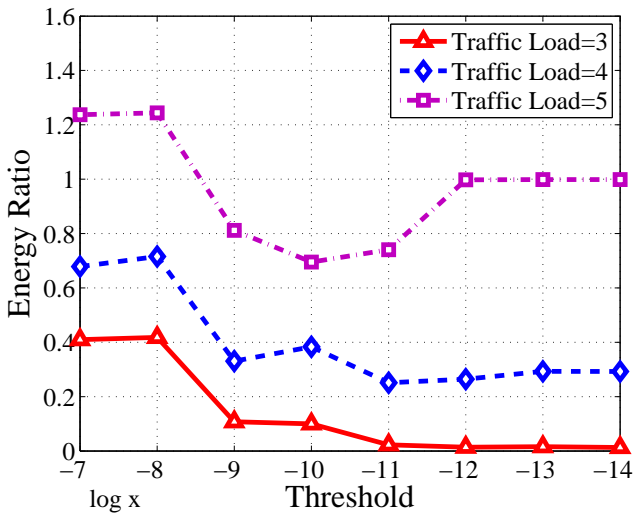}
\end{center}
\caption{The energy ratio of our scheme over TDMA under Traffic A.}
\label{ER_tl}
\end{figure}

In Fig. \ref{ER_tl}, we plot the energy ratio of our scheme over TDMA under different traffic loads. Since the threshold is very small, we show the results with x axis using the logarithmic coordinates. From the results, we can observe that the energy ratio can be achieved lower under light load, which is consistent with the results in Fig. \ref{Energy_A}. The choice of the threshold also has an important impact on the consumed energy. When the traffic load is 5, the energy ratio is still larger than 1 at the threshold of ${10^{ - 8}}$, while the energy ratio is less than 1 when the traffic load is 3 and 4. Thus, the threshold should be smaller to reduce energy consumption when the requested throughput of flows increases, which is consistent with the results in Section \ref{pre-S6}. When the threshold is too small, e.g., ${10^{-12}}$, our scheme will reduce to the serial TDMA scheme, and the energy ratio is 1. However, under light load, more CTAs are allocated to each flow, and the transmission power can also be reduced, which eventually reduces the consumed energy.

\begin{figure} [htbp]
\begin{center}
\includegraphics*[width=9cm]{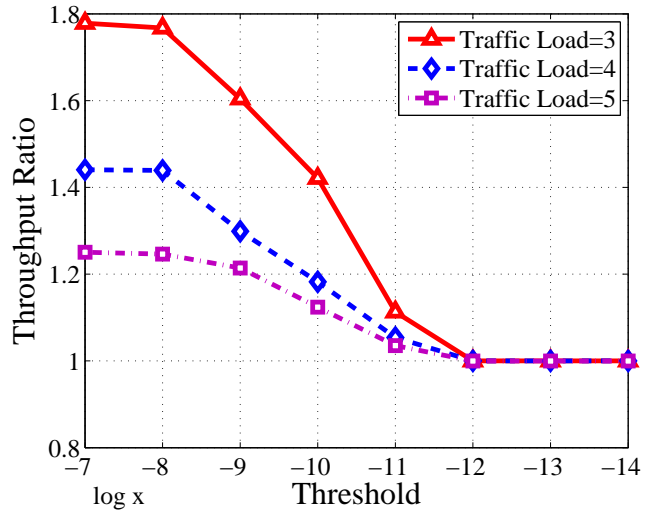}
\end{center}
\caption{The throughput ratio of our scheme over TDMA under Traffic A.}
\label{TR_tl}
\end{figure}

We also plot the throughput ratio of our scheme over TDMA under Traffic A in Fig. \ref{TR_tl}. Our scheme achieves higher throughput ratio under light load. Under light load, there are more CTAs allocated to each flow, and the transmission power of each flow can be reduced further. The actual transmission rate can be larger than the assumed transmission rate, $R_i''$ in (\ref{Assume_R}). Thus, the throughput ratio is higher under light load. When the threshold is as small as ${10^{-12}}$, our scheme reduces to the serial TDMA scheme, and thus the throughput ratio becomes 1.

\subsubsection{Under Different BS Distributions}

\begin{figure} [htbp]
\begin{center}
\includegraphics*[width=9cm]{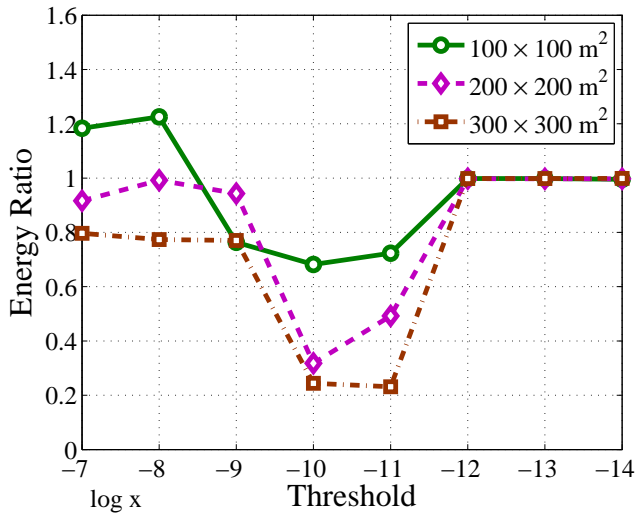}
\end{center}
\caption{The energy ratio of our scheme over TDMA under different BS distributions.}
\label{ER_R}
\end{figure}

In Fig. \ref{ER_R}, we plot the energy ratio of our scheme over TDMA under different BS distributions. We investigate three cases, where the BSs are distributed uniformly in a square area of 100 $ \times$ 100 $m^2$, 200 $ \times$ 200 $m^2$, and 300 $ \times$ 300 $m^2$, respectively. There are ten flows in the network, and the requested throughput of each flow is uniformly distributed between $[2.5, 3.5]$ Gbps. From the results, we can observe that the energy ratio in different BS distributions achieves the different lowest values with different thresholds. Our scheme in the 100 $ \times$ 100 $m^2$ area achieves the energy ratio of about 0.68 with the threshold equal to ${10^{ - 10}}$, and in the 200 $ \times$ 200 $m^2$ area achieves the energy ratio of about 0.32 with the threshold equal to ${10^{ - 10}}$. In the 300 $ \times$ 300 $m^2$ area, our scheme achieves the energy ratio of about 0.23 with the threshold equal to ${10^{ - 11}}$. We can infer that the energy ratio can be achieved lower when the BSs are distributed more sparsely. This is due to in the sparsely distributed case, the interference becomes lower, and spatial reuse can be exploited for energy consumption reduction more efficiently. We only evaluate a finite number of thresholds here, and the threshold should be selected optimally under different BS distributions. If the threshold is too small to enable concurrent transmissions, such as ${10^{ - 12}}$, our scheme will reduce to the serial TDMA scheme, and the energy ratio will be equal to 1. If the threshold is too large, there will be severe interference between concurrent flows, and the number of CTAs allocated to each flow is even smaller than that under the serial TDMA scheme. Thus, the transmission power of flows cannot be reduced significantly, and is even larger than $P_t$, which leads to the energy ratio is close to or larger than 1.

\begin{figure} [htbp]
\begin{center}
\includegraphics*[width=9cm]{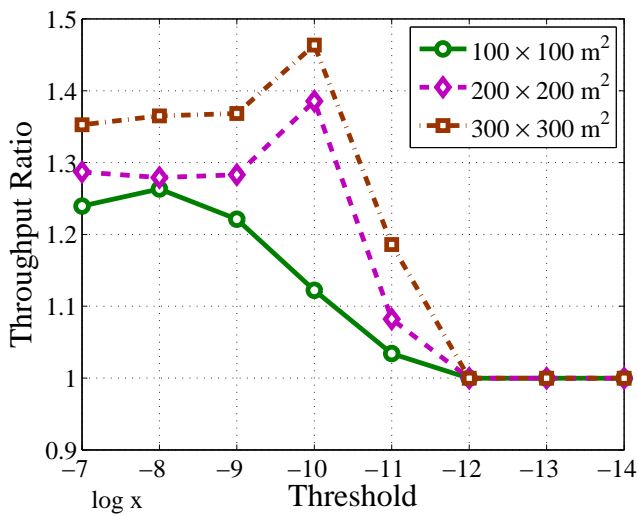}
\end{center}
\caption{The throughput ratio of our scheme over TDMA under different BS distributions.}
\label{TR_R}
\end{figure}

We also plot the throughput ratio of our scheme over TDMA under different BS distributions in Fig. \ref{TR_R}. We can observe that when the threshold is small, such as ${10^{ - 12}}$, concurrent transmissions are not enabled, and the throughput ratio is equal to 1. The maximum throughput ratio is achieved at different thresholds under different BS distributions. In the 100 $ \times$ 100 $m^2$ area, our scheme achieves the throughput ratio of about 1.26 at the threshold of ${10^{ - 8}}$, while our scheme achieves the throughput ratios of about 1.39 and 1.46 at the threshold of ${10^{ - 10}}$ under the 200 $ \times$ 200 $m^2$ area and 300 $ \times$ 300 $m^2$ area, respectively. We can observe that the achieved throughput ratio can be larger when the BSs are distributed more sparsely. The threshold should also be selected elaborately to achieve a high throughput ratio under different BS distributions.


\subsubsection{Under Different Maximum Transmission Power}

\begin{figure} [htbp]
\begin{center}
\includegraphics*[width=9cm]{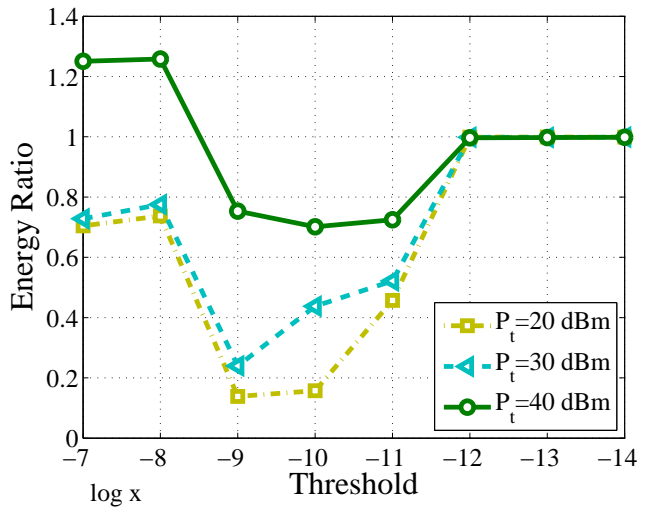}
\end{center}
\caption{The energy ratio of our scheme over TDMA under different maximum transmission power, $P_t$.}
\label{ER_P}
\end{figure}

In Fig. \ref{ER_P}, we plot the energy ratio of our scheme over TDMA under different maximum transmission power, $P_t$. We investigate three cases, where $P_t$ is equal to 20 dBm, 30 dBm, and 40 dBm, respectively. The BSs are distributed uniformly in a square area of 100 $ \times$ 100 $m^2$. There are also ten flows in the network, and the requested throughput of each flow is uniformly distributed between $[2.5, 3.5]$ Gbps. Similar to the results in Fig. \ref{ER_R}, the choice of the threshold is different for our scheme with different $P_t$. We can observe that the optimized threshold to achieve the lowest energy ratio is different with different $P_t$, and the achieved energy ratio is also different. The energy ratio can be lower with the decrease of $P_t$. With lower maximum transmission power, $P_t$, the interference between concurrent flows will be reduced, which benefits exploiting spatial reuse for lower energy consumption.

\begin{figure} [htbp]
\begin{center}
\includegraphics*[width=9cm]{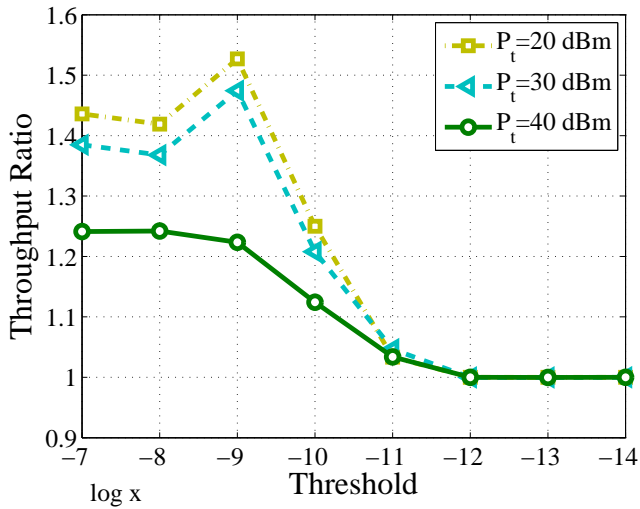}
\end{center}
\caption{The throughput ratio of our scheme over TDMA under different maximum transmission power, $P_t$.}
\label{TR_P}
\end{figure}

We also plot the throughput ratio of our scheme over TDMA under different $P_t$ in Fig. \ref{TR_P}. We can observe that the throughput ratio increases with the decrease of $P_t$. Combing the results in Fig. \ref{ER_P} and Fig. \ref{TR_P}, the threshold should be selected elaborately under different $P_t$ to achieve a low energy ratio or a high throughput ratio.

In summary, the interference threshold should be optimized under different traffic loads, BS distributions, and maximum transmission power to achieve optimal network performance in terms of energy consumption and throughput in practice.

\section{Conclusion}\label{S7} 

In this paper, we investigate the problem of minimizing the energy consumption via optimizing concurrent transmission scheduling and power control for the mmWave backhauling of small cells densely deployed in HCNs. We propose an energy efficient mmWave backhauling scheme, where the concurrent transmissions are exploited for lower energy consumption and higher energy efficiency. The conditions of our scheme to reduce energy consumption by allocating more CTAs to flows and the choice of the interference threshold are also analyzed to provide guidelines for parameter selection in practice. Extensive simulations demonstrate our scheme achieves the lowest energy consumption and highest energy efficiency compared with other two schemes. By investigating the performance of our scheme with different interference thresholds, we show the threshold should be selected elaborately under different traffic loads, BS distributions, and the maximum transmission power to achieve low energy consumption and high energy efficiency.

\end{document}